# Plastic strain accommodation and acoustic emission during melting of embedded particles

Michael Kuba[a] and David C. Van Aken[a,+]

[a]223 McNutt Hall, 1400 N. Bishop, Rolla, MO 65409-0330, USA

[+]Corresponding Author. dcva@mst.edu

Melting point phenomena of micron-sized indium particles embedded in an aluminum matrix were studied by means of acoustic emission. The acoustic energy measured during melting increased with indium content. Acoustic emission during the melting transformation suggests a dislocation generation mechanism to accommodate the 2.5% volume strain required for melting of the embedded particles. A geometrically necessary increase in dislocation density of 4.1 x 10$^{13}$ m$^{-2}$ was calculated for the 17 wt% indium composition.

Embedding small particles (micron-sized) in a higher melting point matrix is known to increase the melting temperature of the particles. Numerous causes for the phenomenon have been proposed, including strain energy effects [1,2], interfacial energy effects [3-5], and kinetic barriers to nucleation [6].

Malhotra and Van Aken studied the anelastic strain accommodation during melting of micron-sized indium particles embedded in an aluminum matrix [7-9]. By measuring internal friction and performing differential scanning calorimetry (DSC), they concluded that the noted increase in melting temperature was mainly a strain energy effect from the volume expansion during the melting transformation.

Acoustic emission (AE) describes the propagation of elastic waves resulting from rapid energy release in a material [10]. Two qualitative types of AE exist: "burst," a discrete signal, and "continuous," a sustained signal usually caused by several bursts overlapping [10]. For example, crack growth tends to generate a burst emission, while dislocation movements result in a continuous emission [11]. Phase transformations that generate AE usually exhibit continuous emission due to time dependent nucleation [11].

According to literature, displacive solid-state transformations exhibit AE; the shear mechanism results in rapid strain energy release in the form of AE. Diffusive transformations occur too slowly for this effect [12]. In steels, formation of allotriomorphic ferrite or pearlite would not generate AE [12], but martensite [12] and bainite [13] do. Widmanstätten ferrite may also exhibit AE [13]. AE is often recommended for use as a criterion in determining the displacive or martensitic-like qualities of a solid-state transformation [14]. However, solid-liquid transformations also exhibit AE as the material shrinks [15], e.g. indium would only exhibit AE upon solidification. The exact cause of AE during melting and solidification is controversial [16], but may be due to frictional noise between solid crystals [17] or cluster addition/subtraction from the solid-liquid interface [18]. In polymers, AE during crystallization is due to cavitation in areas of occluded liquid where shrinkage stresses overwhelm the cohesive strength of the melt [19].

AE techniques also detect other phenomena involving dislocation creation and movement [20], yield point deformation, and fracture processes during tensile testing [21,22]. It is suggested here that the stress conditions around an indium particle during melting are similar to that required for void nucleation and growth during ductile fracture, i.e. the volume increase during melting may also be accommodated by dislocation generation and that this process may be reversible. Consequently, it was hypothesized that the melting reaction might generate AE.

The scope of this article is to show successful detection of the melting of indium particles embedded in an aluminum matrix by AE. By investigating the nature of the reaction, a better understanding of solid-state transformations may be realized.

Bulk material for testing was prepared by melting aluminum in fireclay crucibles in a resistance furnace at 800°C. Indium pieces were wrapped in aluminum and plunged into each melt to create a range of aluminum-indium alloys with nominal chemistries of 0, 4, 8, 12, and 17 wt% indium. The melt was physically stirred to aid in homogenization of the melt since liquid miscibility is possible at 800°C. All materials used were at least 99.99% pure with respect to metal content. The alloys were chill cast into 13 mm diameter cylinders using an aluminum mold. The microstructure was characterized using a Hitachi S-570 scanning electron microscope. Secondary electron images were obtained and exhibit atomic weight contrast due to backscattered electrons interacting with the pole piece of the microscope. Representative micrographs are shown in Figure 1. The Al-17In composition is near the monotectic composition and produced a uniform distribution of micron-diameter indium particles. Decreasing the indium content produced an increasingly dendritic microstructure with indium particles residing in interdendritic regions. The size of the indium particles also increased with decreasing indium content. Three images of different areas for each composition were analyzed using ImageJ software to determine volume fraction of indium, which was then converted to weight percentage using appropriate densities. Compositional results are shown in Table 1.

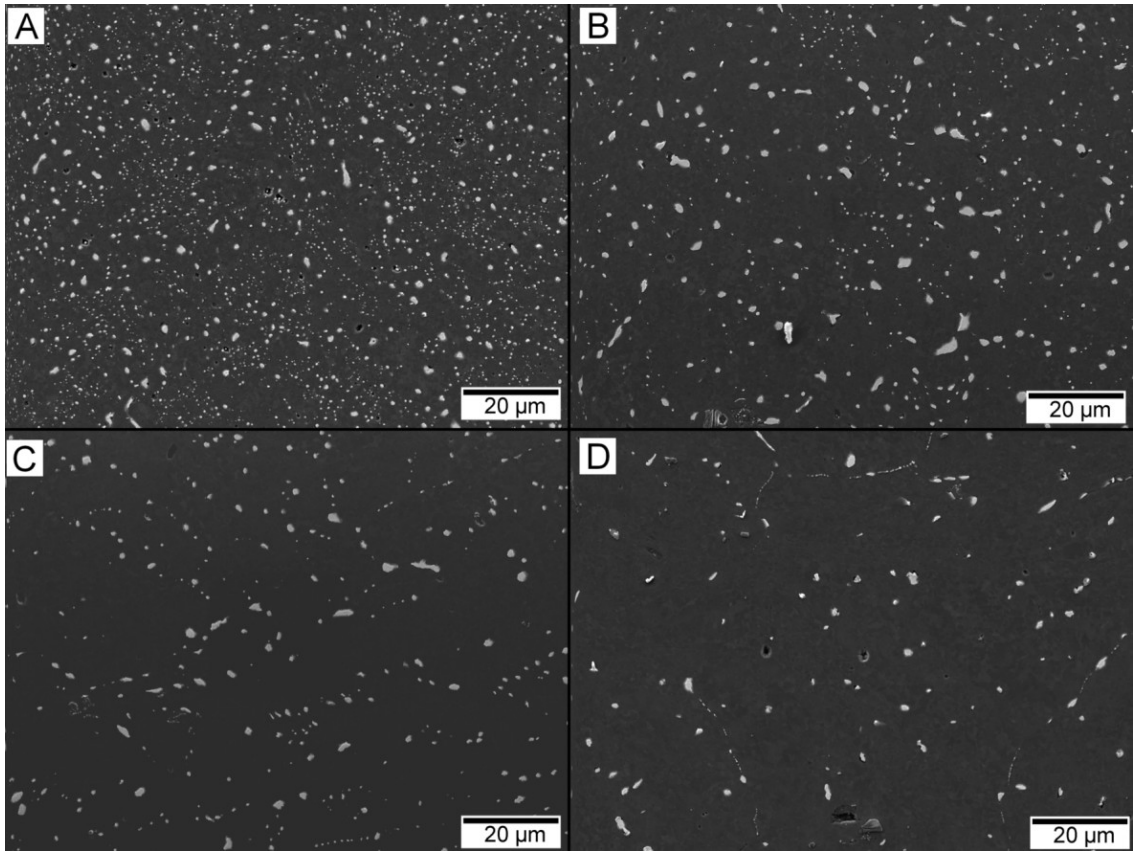

Figure 1. Secondary electron images of the four Al-In compositions studied. Micrographs show (a) 17 wt% In, (b) 12 wt% In, (c) 8 wt% In, and (d) 4 wt% In.

DSC was performed to confirm desired microstructural characteristics, i.e. the elevated melting point. Results for each of the cast alloys were similar to that reported by Malhotra and Van Aken [7]. Heat flow curves had a sharp peak initiating near 155.5°C, terminating near 160°C, and peaking near 157°C; while a broader peak was also present between 155°C and 170°C, which peaked between 163°C and 166°C. The two peaks were deconvoluted by assuming superimposed Gaussian distributions. The first peak is interpreted as indium particles melting at the equilibrium temperature, 156.6°C [23], while the second peak represents elevated temperature melting as investigated previously by Malhotra and Van Aken [7]. The equilibrium peak ranged between 17% and 36% of the overall reaction, while the elevated melting peak ranged between 64% and 83%.

AE testing was performed on as-cast specimens, machined into right cylinders with three orthogonal holes drilled normal to the surfaces, which produced a wall thickness of 4 mm. The three holes were drilled to reduce thermal gradients in the specimen. Specimen temperature was recorded from a thermocouple swaged into a 2.25 x 2.25 mm hole with machining chips of the same composition. Specimens were weighed before testing to relate acoustic energy with volume of indium transformed.

A high temperature epoxy was used to attach the specimens to a 12.7 mm diameter aluminum alloy 6061 waveguide. A cross beam was mounted to the wave guide to suspend the specimen into a molten salt bath held at 200°C. A PZT Navy type V transducer produced by Physical Acoustics Corporation was clamped to the fixture with Dow Corning high vacuum grease as a couplant. AE and specimen temperature were monitored using National Instruments LabVIEW software. The specimen heating rate was measured to be 0.16 to 0.29 °C/s.

AE was monitored by an average signal level measured in decibels with a time constant of 0.1 seconds, rather than by hit count, as used by Van Bohemen [24]. The average signal level was related to the voltage level using Equation 1, where $V_{RMS}$ is the root mean square signal voltage in microvolts at the preamplifier input and $dB_{AE}$ is the signal amplitude in decibels. A sinusoidal waveform was assumed when relating the average signal level to the RMS voltage. The transducer signal was passed through a 40 dB gain preamplifier with a threshold of 25 dB and a software bandpass filter from 1-3000 kHz.

Table 1. The area percentage analysis and converted weight percentage of the cast alloys.

| Nominal Composition | Area Percentage Indium | Weight Percentage Indium | Average Particle Volume (μm³) |
|---|---|---|---|
| Al-17In | 5.8 | 15.70 | 0.165 |
| Al-12In | 3.9 | 10.56 | 0.465 |
| Al-8In | 2.1 | 5.69 | 0.425 |
| Al-4In | 1.4 | 3.79 | 0.659 |

Equation 1 $$V_{RMS} = \frac{10^{(dB_{AE})/20}}{\sqrt{2}}$$

Typical plots of AE versus temperature for each alloy are shown in Figure 2. The pure aluminum specimen does not exhibit AE in the temperature range of interest. However, the aluminum-indium alloys do exhibit AE and the AE increases with increasing indium content. Specimens exhibited a melting temperature ranging from 157 to 165°C based on temperature measurements during AE. Differences with DSC measurements are ascribed to inconsistencies in the thermocouples used and varying specimen thermal lag.

The energy of transformation is directly proportional to the integral of the square of the RMS voltage of the AE signal with respect to time during the transformation [25]. The data show an increasing trend of transformational energy with increased volume of indium as shown in Figure 3.

Prismatic dislocation loop emission is proposed for the volume accommodation of the melting indium particles. Molecular dynamics simulation work and spall testing by D. C. Ahn et al. [26,27] and laser shock tests by Lubarda et al. [28] confirm dislocation loop emission as a viable mechanism for void growth at temperatures too low for diffusion to occur within the necessary time scale.

Average particle volume as a function of composition is listed in Table 1. Malhotra and Van Aken estimated a volume change of 2.5% upon melting [7,8]. Following the analysis of Ahn et al. [26], the volume change due to the emission of multiple prismatic loops can be related to the number of emitted dislocations by Equation 2. Here ΔV is the change in volume, N is the number of dislocation loops, $\rho_0$ is the loop radius (assumed to be 75% of the particle radius [26]), b is the Burgers vector magnitude, and $\sum_{i=1}^{N} \Delta V^*(z_i)$ accounts for the elastic volume change due to the loops being in the vicinity of the void. For voids larger than 400 times the Burgers vector, the elastic volume change, dependence on void radius, image stresses, and surfaces energy effects can be ignored [26]. Thus, the emission of 115, 100, 103, and 73 prismatic dislocation loops are expected per indium particle for the 4, 8, 12, and 17 wt% indium alloys.

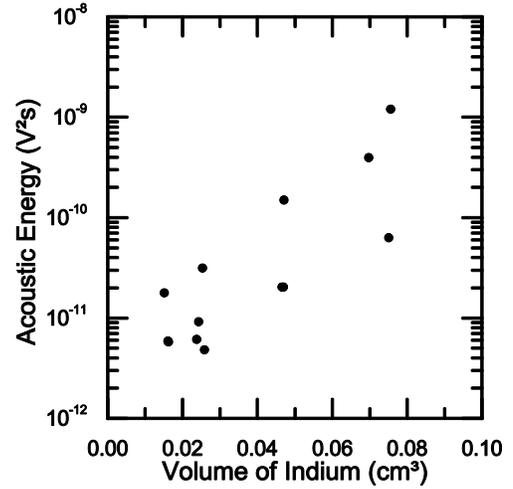

Figure 3. The measured acoustic energy related to the volume of indium particles in each specimen.

Equation 2 $$\Delta V = N\pi\rho_0^2 b + \sum_{i=1}^{N} \Delta V^*(z_i)$$

The total change in dislocation density (geometrically necessary) can now be calculated by estimating the number of average particles necessary to account for the measured volume fraction of indium in each alloy. The total dislocation density would increase by $6.3 \times 10^{12}$, $1.1 \times 10^{13}$, $2.0 \times 10^{13}$, and $4.1 \times 10^{13}$ m$^{-2}$ for the 4, 8, 12, and 17 wt% indium alloys. For comparison, it is interesting to note that Widmanstätten ferrite formation generates a dislocation density of $2.58 \times 10^{14}$ m$^{-2}$ [29]. As well, bainite is expected to generate a dislocation density on the order of $1.7 \times 10^{14}$ m$^{-2}$ for continuously cooled steel [30] and $4 \times 10^{14}$ m$^{-2}$ for steel isothermally transformed at 650°C [31]. It is estimated that the acoustic energy generated during melting is approximately 25% of that reported by Van Bohemen et al. [13] for the bainitic reaction, after normalizing for the volume transformed. This is comparable to the difference between dislocation densities calculated for the Al-17In melting reaction and that observed in bainite. Thus, AE detection of bainite can be explained by a mechanism of volume strain accommodation by dislocation generation, rather than a displacive transformation mechanism.

The model of prismatic loop emission for volume accommodation can also be correlated to internal friction results from Malhotra and Van Aken [8] and Wolfenden and Robinson [32]. Both studies noted that the nearly-immiscible second phase particles (pure lead and pure indium) sit at both grain boundaries and within matrix grains. Malhotra and Van Aken's internal friction and DSC results were interpreted as a diffusional relaxation with particles sitting at grain boundaries having a much shorter relaxation time than those embedded in the grain interiors. A dislocation generation model had not been previously considered, but the AE shown in this study strongly suggests that dislocations are generated. As grain boundaries are an easy source for dislocation generation [33], the time for grain boundary relaxation is expected to be significantly shorter than a diffusional relaxation mechanism. It is well known that internal friction results are a maximum when the product of the test frequency and the relaxation time approach unity. Consequently, the lower test frequencies (0.5 – 1 Hz) used by Malhotra and Van Aken would emphasize a diffusional relaxation model, whereas the 40 kHz test frequency used by Wolfenden and Robinson may emphasize a much shorter relaxation time typical of dislocation generation.

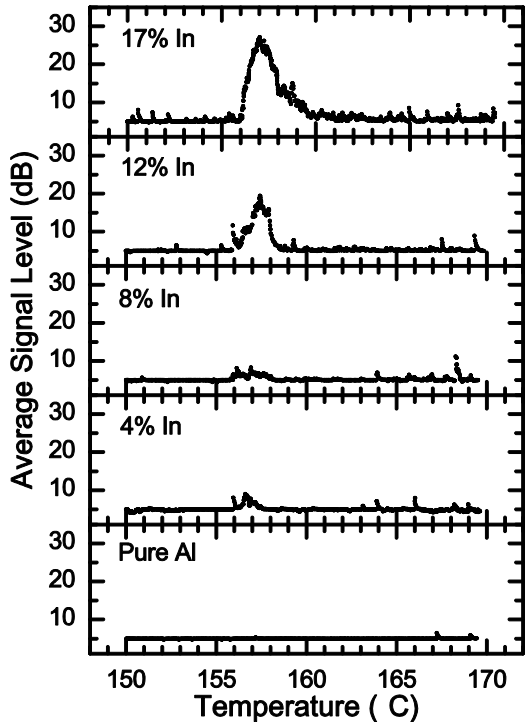

Figure 2. Average signal level detected versus temperature for each composition studied. Thermal lag was removed and initiation of the AE peak was placed at 156°C. A logarithmic smoothing algorithm was applied.

In summary, aluminum-indium alloys were shown to generate AE during melting of the embedded indium particles. A dislocation mechanism for the volume strain accommodation was proposed and the dislocation density calculated seems reasonable with respect to the measured AE. In a similar fashion, a dislocation mechanism for volume strain accommodation may also explain the high dislocation densities reported for bainitic transformations without evoking a displacive transformation mechanism. Furthermore, it may be speculated that the easy nucleation of dislocations from grain boundaries may be more important than surface energy effects in causing preferential nucleation at grain boundaries during solid-state phase transformations.

This work was supported in part by the National Science Foundation, the Department of Energy, and the American Iron and Steel Institute under contract No. CMMI 0726888.